# Magnetic transitions in the 1-D chain compounds NdPd$_5$Ge$_3$ and NdPt$_5$Ge$_3$

*Xin Gui and Robert J. Cava*[*]

Department of Chemistry, Princeton University, Princeton NJ 08540, USA

### ABSTRACT

We present the structural and initial magnetic characterization of the previously unreported materials NdPd$_5$Ge$_3$ and NdPd$_5$Ge$_3$. These materials, with 1-dimensional Nd chains, crystallize in the orthorhombic YNi$_5$Si$_3$-type structure in space group *Pnma*. Magnetic ordering is observed for both compounds at ~ 2.2 K for NdPd$_5$Ge$_3$ and ~ 3.0 K for NdPt$_5$Ge$_3$. A magnetic-field-induced transition is clearly observed for NdPd$_5$Ge$_3$ below 2K, at an applied magnetic field of around 1.6 Tesla. The heat capacity data reveals that essentially all the available magnetic entropy is released at the magnetic ordering transition for NdPd$_5$Ge$_3$ but that only 29% is seen for NdPt$_5$Ge$_3$ at temperatures down to 1.5 K, the data implying that a second magnetic transition can be expected at lower temperatures.

*Address correspondence to: rcava@princeton.edu

## *1. Introduction*

Rare-earth-based intermetallics have attracted interest due in large part to their magnetic properties, which often reflect different degrees of hybridization between the rare earth $f$ electrons and the other atomic orbitals present. Intriguing electronic and magnetic properties, such as heavy-fermion superconductivity, long-range magnetic ordering, and Kondo insulator behavior have been observed in such systems.[1–9] In addition, materials containing heavy elements with strong spin-orbit coupling, like Pd and Pt, are of interest due to such atom's ability polarize, split or invert energy bands, such as has been observed in topological materials.[10–12]

Motivated by these considerations, we here report the synthesis, structure, and magnetic characterization of $NdPd_5Ge_3$ and $NdPt_5Ge_3$. The Cerium analogs, $CePd_5Ge_3$ and $CePt_5Ge_3$, which are isostructural, antiferromagnetically order (1.9 K for $CePd_5Ge_3$ and 1.1 K for $CePt_5Ge_3$) with possible low-dimensional magnetism observed for the latter[13–16]. For $CePt_5Ge_3$, a tail in the specific heat was observed, indicating antiferromagnetic short-range correlations of a low-dimensional character above the 3-D long-range order.[16] Moreover, magnetic-field-induced magnetic transitions were observed in $CePd_5Ge_3$ - two successive transitions, a spin-flop transition and an AFM to paramagnetic state transition can be seen when magnetic field is applied along a direction close to $b$ axis.[15] We find that the Nd-based materials show slightly higher magnetic ordering temperatures than do their less strongly magnetic Ce analogs and other differences in the magnetism. Of interest from the structural viewpoint in these materials are the well separated 1-dimensional (1-D) chains made from the rare-earth atoms. Our magnetic characterization shows that the two new materials studied have similar magnetic properties, and suggest that those properties mmay be very sensitive to the spin-orbit coupling differences between Pd and Pt.

## *2. Experiment*

The arc melting method, under a high purity Zr-gettered, argon atmosphere, was employed to synthesize polycrystalline samples of both materials. Elemental palladium (99.95%, APS 0.5-1.7 micron, Alfa Aesar) and platinum (>99.98%, ~60 mesh, Alfa Aesar) were first pressed into pellets and arc melted. PdGe and PtGe precursors were then made by arc melting the pre-melted Pd/Pt chunks with germanium (≥99.9999%, pieces, Alfa Aesar) in a molar ratio of 1:1.02 to compensate for the mass loss of Ge. Finally, the ternary $NdM_5Ge_3$ materials were synthesized by mixing elemental Nd (≥99%, ingot, Beantown Chemical), the PdGe/PtGe precursors and pre-melted Pd or Pt in the molar ratio of 1:3:2.1. The mixture was arc melted three times with the melted buttons turned over between melts. The final products are air-stable.

A Bruker D8 Advance Eco diffractometer with Cu Kα radiation and a LynxEye-XE detector was employed to obtain the powder X-ray diffraction (PXRD) patterns for both materials at ambient temperature. The patterns were fitted by the Rietveld method in *Fullprof* using as an initial guess the crystal structure obtained from the Inorganic Crystal Structure Database (ICSD) for $CePd_5Ge_3$.[13]

The DC magnetic susceptibility data were obtained by utilizing a Dynacool Physical Property Measurement System (PPMS, Quantum Design Inc.) in the temperature range of 1.8 K to 300 K, and, when necessary, under different applied magnetic fields. The magnetic susceptibility was defined as M/H where H is the applied magnetic field in Oe and M is the measured magnetization in emu. Heat capacity was measured using a standard relaxation method in the PPMS between 1.5 K to 10 K with the $^3$He function.

## 3. Results

**3.1. Crystal Structure:** The powder XRD patterns of $NdPd_5Ge_3$ and $NdPt_5Ge_3$ fit well to the model $CePd_5Ge_3$ structure, as shown in Figures 1(a) and (b). The materials obtained have high purity. The Rietveld fitting parameters and refined atomic positions are listed in Tables 1 and 2. The materials are isostructural with $CePd_5Ge_3$,[13] which crystallizes in the orthorhombic $YNi_5Si_3$-type structure with space group *Pnma* (No. 62). In our materials, replacing the Pd with Pt increases $a$, $b$ and $c$ by 0.3%, 0.5% and 0.2%, respectively.

As is shown in Figure 2(a), the structure consists of rare-earth-based $Nd@Ge_6$ prisms imbedded in a framework of Pd or Pt-based $M@Ge_4$ tetrahedra and $M@Ge_5$ pyramids. The Nd atoms form 1-dimensional (1-D) chains along the crystallographic $b$ axis. The coordination polyhedra are shown in Figure 2(b). The symmetry equivalent chains in the structure are arranged along the $a$ axis in a staggered fashion. In Figure 2(c), which shows the Nd atoms only, the Nd-Nd distances are marked as $d_1$, $d_2$ and $d_3$, which are the shortest Nd-Nd bond lengths within and between the different Nd chains. For both compounds, $d_1$ is the same as $d_2$ while $d_3$ is larger. As shown in Table 3, the Nd-Nd distance within the chains is 0.5% longer in $NdPd_5Ge_3$ than in $NdPt_5Ge_3$.

**3.2. Magnetic properties:**

**3.2.1 NdPd₅Ge₃:** As illustrated in Figure 3(a), under an applied external field of 0.3 T, Curie-Weiss behavior is found for $NdPd_5Ge_3$ in the temperature range of 25 K to 300 K. The data can be fit with the Curie-Weiss (CW) law: $\chi = C/(T-\theta_{CW})$ where $\chi$ is the measured magnetic susceptibility of the material, C is independent of temperature, related to the effective moment, and $\theta_{CW}$ is the Curie-Weiss theta (Temperature independent contributions to the magnetic susceptibility are not significant in magnitude in these materials.) In $NdPd_5Ge_3$, $\theta_{CW}$ is -6.41 (1) K and the effective moment $\mu_{eff} = \sqrt{8C}$ is ~3.71 $\mu_B$/f.u., which is close to the expected effective moment of the free $Nd^{3+}$ ion (3.62 $\mu_B$). To better visualize the low-temperature magnetic behavior of this material, $\chi$ *vs.* T curves under an applied field of 50 Oe, measured in both the zero-field cooling (ZFC) and field cooling (FC) modes, are presented in Figure 3(b). The shape of the FC curve at low temperature suggests ferromagnetic ordering. A drop of susceptibility was found at ~ 2.2 K for the ZFC curve which we attribute to the movement of magnetic domain walls. The inset of Figure 3(b) presents the first derivative of the $\chi$ *vs.* T curve ($d\chi/dT$ for the ZFC mode), which reveals that non-CW behavior starts from ~ 3.5 K, as indicated by the divergence of the $d\chi/dT$ curve from $d\chi/dT = 0$.

Figures 3(c) and (d) show the hysteresis loops for $NdPd_5Ge_3$ in various field ranges at various temperatures. In the low-field region (-0.05 T to 0.05 T), the magnetization for this material reflects the presence of a small ferromagnetic (FM) component (~ 0.39 $\mu_B$/$Nd^{3+}$ at 1.8 K) at temperatures below 3 K. A coercive field around of 20 Oe can be seen at 1.8 K, indicating FM ordering. Above 3 K, the FM contribution vanishes, which reveals that the moments in $NdPd_5Ge_3$ are not aligned ferromagnetically above 3 K. Thus, the observed low field FM component is a consequence of the ordered state in $NdPd_5Ge_3$ instead of the presence of trace elemental FM impurities. Below $T_M$, a magnetic-field-induced transition is observed at ~ 1.6 T. In the inset of Figure 3(c), we enlarge the magnetization curve from 1 T to 3 T at 1.8 K and 2 K to illustrate this. After the spontaneous magnetization at low field (<< 1 T), the magnetization curves at both 1.8 K and 2 K continue to increase in a nearly linear way until the magnetic field of ~1.6 T, where both curves undergo a transition to different slopes, i.e., a different magnetic state. As can be seen in inset of Figure 3(d), broad peaks indicating the likely presence of a field-induced transition to a different magnetic state are observed in the first derivative curves between 1 and 3.5 T at 1.8 K and 2 K, while at 2.5 K, no such transition is observed up to fields of 7 Tesla. At higher applied fields, the magnetization of $NdPd_5Ge_3$ saturates at 1.8 K with moment at 9 T M ~ 1.93 $\mu_B$/$Nd^{3+}$, which is smaller than the saturated moment for $Nd^{3+}$. Thus, by 9 T, three $f$ electrons on Nd are not totally aligned by the field.

**3.2.2 NdPt₅Ge₃:** The temperature-dependent magnetic susceptibility for $NdPt_5Ge_3$ can be seen in Figure 4(a). Curie-Weiss behavior is also observed in this case. The CW temperature from the Curie-Weiss fit, $\theta_{CW}$, is -5.69 (4) K, while the effective moment is ~3.76 $\mu_B$ per Nd atom, similar to what is observed for the Pd variant. When the external magnetic field is smaller, i.e., 50 Oe, $NdPt_5Ge_3$ exhibits a broader magnetic ordering transition than is seen for $NdPd_5Ge_3$, at ~ 3.0 K. The inset of Figure 4(b) shows the first derivative of $\chi$ $vs$ T measured in the ZFC mode between 2 and 6 K; the nonlinear (i.e. non-CW) behavior starts from ~4.9 K.

The hysteresis loops shown in Figure 4(c) and (d) reveal a magnetic moment M ~ 1.81 $\mu_B$/$Nd^{3+}$ at 1.8 K and 9 T. Magnetic hysteresis was observed at 4 K and below, indicating FM ordering. Similar with $NdPd_5Ge_3$, three $f$ electrons on Nd are not totally aligned by the field by 9 T. The coercive field can be extracted as about 110 Oe at 1.8 K while it is around 6.5 Oe at 4 K.

**3.3. Heat Capacity:** Figure 5 describes the heat capacity behavior of both compounds from 1.5 K to 10 K. The total heat capacity of a magnetic material can be interpreted as the sum of electronic plus phonon plus magnetic contributions, as for $C_{total} = C_{elect} + C_{phonon} + C_{mag}$. Due to the lack of non-magnetic model of such system, the electronic contribution could be described as $C_{elect} = \gamma T$ where g is the Sommerfeld coefficient while the phonon contribution could be estimated by using $C_{phonon} = \beta T^3$ according to Debye formula where $\beta$ is the electronic contribution coefficient.[17] However, the $T^3$ relation cannot lead to a good fitting, thus a $T^5$ term was added. Therefore, the non-magnetic contributions can be estimated by using a polynomial fit to the temperatures above the magnetic transition: $C_{electron+phonon}/T = \gamma + \beta_1 T^2 + \beta_2 T^4$ where a, $b_1$ and $b_2$ are constants and were fitted to 1.18 (2), -0.015 (1), 0.00010 (1) for $NdPd_5Ge_3$ and 0.036 (6), 0.0085 (2), -0.000021 (2) for $NdPt_5Ge_3$. After extraction of the approximate electronic and phononic contributions, the residual heat capacity can be attributed to the magnetic system only. As shown in Figure 5(a), a sharp peak can be found for $NdPd_5Ge_3$ at around 2.5 K and that the entropy does not saturate until 3.5 K. Meanwhile, the entropy change, $\Delta S_{magnon}$, exceeds the value of Rln2 at 2.6 K while reveals that most entropy change from magnetic ordering can be found below 2.6 K, i.e., the system is fully magnetic ordered by that temperature. However, in Figure 5(b), even though a low broad peak is observed at around 3 K, the total entropy change is only ~29% of Rln2, which indicates that most of the entropy from magnetic ordering in $NdPt_5Ge_3$ cannot be seen above 1.5 K, i.e., the magnetic system is not fully ordered down to that temperature. Considering the fact that the total heat capacity of NdPt$_5$Ge$_3$ increases dramatically by 1.5 K, a large magnetic entropy change may be observed below 1.5 K.

## 4. Conclusion

We have synthesized two previously unreported magnetic Nd-based materials, NdPd$_5$Ge$_3$ and NdPt$_5$Ge$_3$. The phases are pure and their crystal structures, determined by powder X-ray diffraction, are very similar to that of CePd$_5$Ge$_3$,[13] suggesting that this structure is stable for germanides for at least the larger rare earths. Initial magnetic characterization of both materials shows magnetic ordering at 3 K or below, with field-induced magnetic state transitions for NdPd$_5$Ge$_3$ at field of ~1.6 T. Heat capacity data shows sharp magnetic ordering with most entropy change appearing above 1.5 K for NdPd$_5$Ge$_3$. However, even though a broad peak is found in the heat capacity at ~3 K for NdPt$_5$Ge$_3$, the entropy sum is only ~23% of Rln2, indicating that by 1.5 K, NdPt$_5$Ge$_3$ does not show complete magnetic ordering. Our data reveal that the magnetic structures of these phases are rather complex, and thus neutron diffraction study would be needed for full characterization.

## Acknowledgements

This work was supported by the US Department of Energy, Division of Basic Energy Sciences, grant DE-FG02-98ER45706.

**Table 1.** Rietveld fitting parameters from the ambient temperature powder Xray diffraction patterns of NdPd$_5$Ge$_3$ and NdPt$_5$Ge$_3$ space group Pnma (No. 62).

|  | R$_p$ (%) | R$_{wp}$ (%) | R$_{exp}$ (%) | χ$^2$ |
|---|---|---|---|---|
| **NdPd$_5$Ge$_3$** | 6.67 | 9.05 | 7.13 | 1.61 |
| **NdPt$_5$Ge$_3$** | 6.63 | 8.80 | 6.40 | 1.89 |

**Table 2.** Structural parameters determined at ambient temperature for NdPd$_5$Ge$_3$ and NdPt$_5$Ge$_3$. In space group Pnma (No. 62) with cell parameters a = 20.1788 (4) Å, b = 4.1066 (1) Å, c = 7.2555 (1) Å for NdPd$_5$Ge$_3$ and a = 20.2457 (3) Å, b = 4.1289 (1) Å, c = 7.2686 (1) Å for NdPt$_5$Ge$_3$.

NdPd$_5$Ge$_3$:

|  | x | y | z |
|---|---|---|---|
| **Nd1** | 0.1431 (2) | ¼ | 0.876 (1) |
| **Pd1** | 0.2958 (3) | ¼ | 0.684 (1) |
| **Pd2** | 0.4961 (2) | ¼ | 0.374 (1) |
| **Pd3** | 0.0152 (3) | ¼ | 0.622 (1) |
| **Pd4** | 0.1147 (2) | ¼ | 0.364 (1) |
| **Pd5** | 0.2949 (4) | ¼ | 0.069 (1) |
| **Ge1** | 0.4174 (4) | ¼ | 0.109 (2) |
| **Ge2** | 0.2386 (4) | ¼ | 0.385 (2) |
| **Ge3** | 0.4176 (3) | ¼ | 0.662 (1) |

NdPt$_5$Ge$_3$:

|  | x | y | z |
|---|---|---|---|
| **Nd1** | 0.1446 (4) | ¼ | 0.881 (1) |
| **Pt1** | 0.2986 (3) | ¼ | 0.681 (1) |
| **Pt2** | 0.4911 (2) | ¼ | 0.365 (1) |
| **Pt3** | 0.0211 (3) | ¼ | 0.622 (1) |
| **Pt4** | 0.1180 (3) | ¼ | 0.354 (1) |
| **Pt5** | 0.2954 (3) | ¼ | 0.090 (1) |
| **Ge1** | 0.415 (1) | ¼ | 0.104 (2) |
| **Ge2** | 0.2508 (6) | ¼ | 0.363 (2) |
| **Ge3** | 0.418 (1) | ¼ | 0.580 (1) |

**Table 3.** Nd-Nd distances for NdPd$_5$Ge$_3$ and NdPt$_5$Ge$_3$ at ambient temperature.

|  | *d1/d2* (Å) | *d3* (Å) | Intrachain distance (Å) |
|---|---|---|---|
| **NdPd$_5$Ge$_3$** | 6.0003 (4) | 6.3856 (4) | 4.107 (1) |
| **NdPt$_5$Ge$_3$** | 5.9730 (4) | 6.4469 (4) | 4.129 (1) |

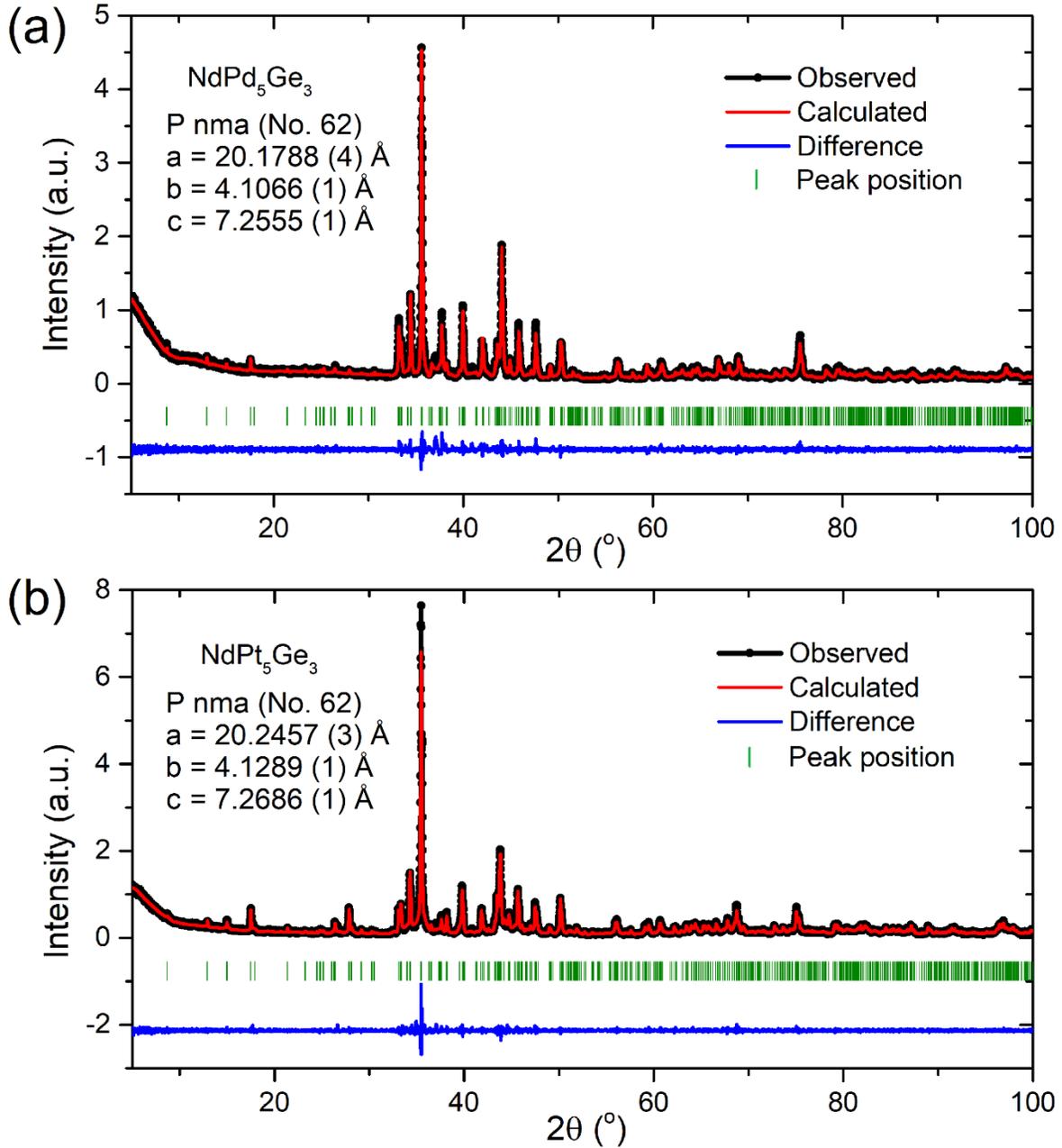

**Figure 1.** Rietveld fitting of the PXRD patterns for **(a)** NdPd$_5$Ge$_3$ and **(b)** NdPt$_5$Ge$_3$. Cu K$\alpha$ radiation, ambient temperature. The numbers in parentheses for the lattice parameters *a b* and *c* indicate the uncertainties obtained in the least-squares fits.

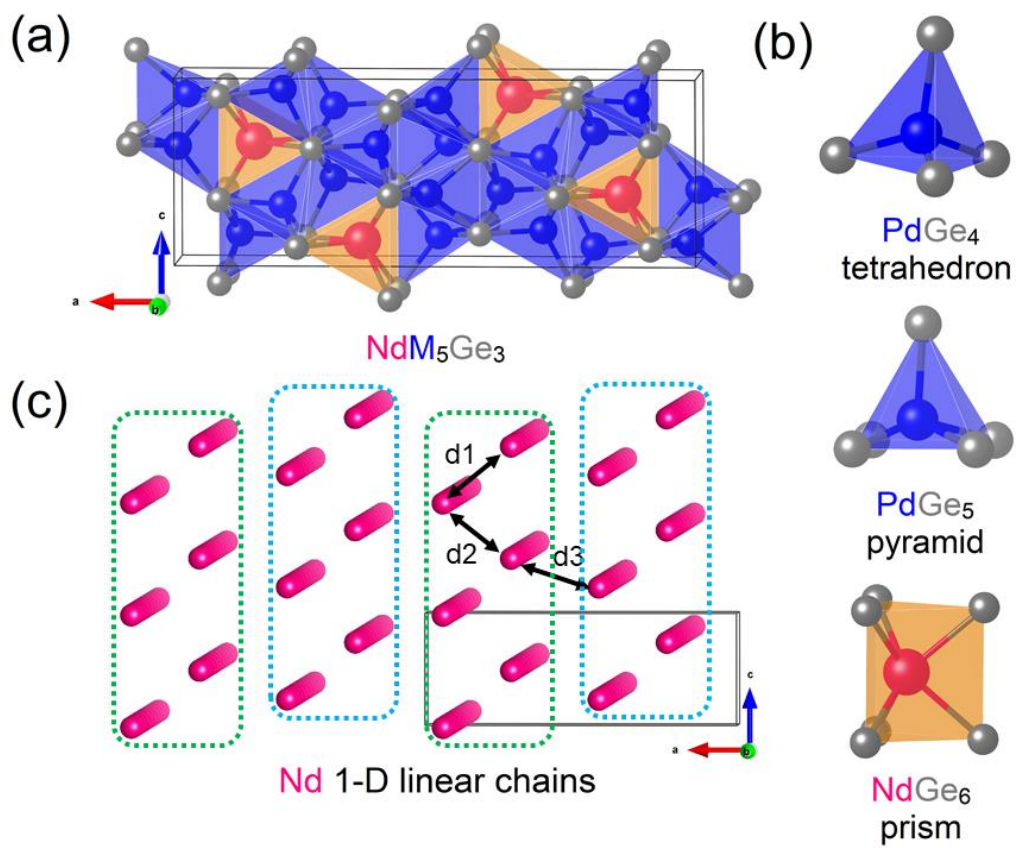

**Figure 2.** (a) The Crystal structure of $NdPd_5Ge_3$ and $NdPd_5Ge_3$. Red, blue and grey balls represent Nd, Pd or Pt (M) and Ge atoms. (b) The $M@Ge_n$ (n = 4 or 5) polyhedra and $Nd@Ge_6$ prisms in the structure. (c) the Nd chains looking down the *b* axis (from a slightly skewed angle to facilitate visualization of the chains.) The symbols *d*1, *d*2 and *d*3 are the shortest distances between the Nd chains within the *ac*-plane.

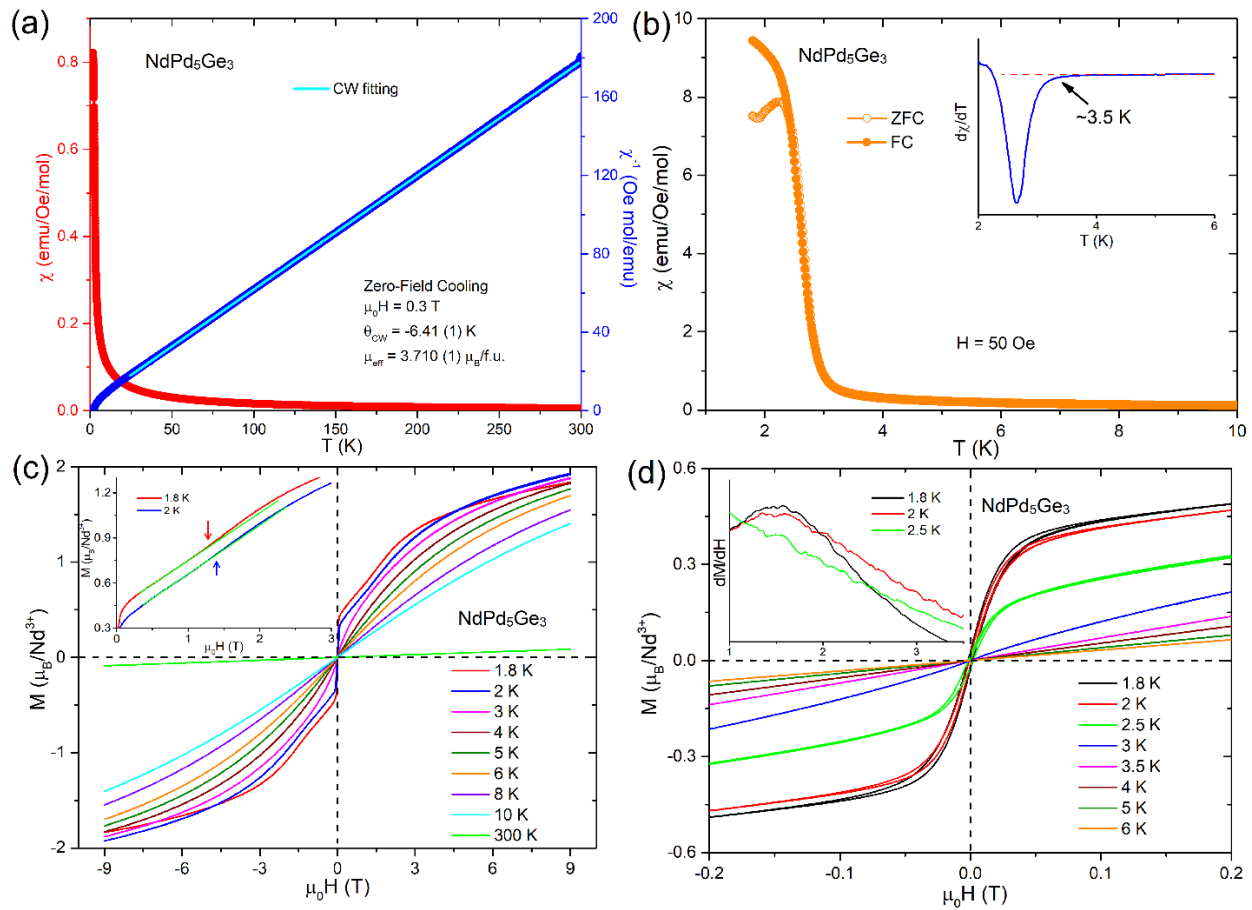

**Figure 3.** **(a)** Temperature-dependence of the observed magnetic susceptibility (χ) and 1/χ of NdPd$_5$Ge$_3$ between 1.8 K and 300 K under applied magnetic field of 0.3 T measured in the ZFC mode. The cyan line represents the Curie-Weiss fitting. **(b)** **(Main panel)** Temperature-dependence of the magnetic susceptibility (χ) of NdPd$_5$Ge$_3$ between 1.8 K and 10 K under an applied magnetic field of 50 Oe (0.005 T). **(Inset)** First derivative of χ vs T measured in the ZFC mode, between 2 K and 6 K. **(c)** **(Main panel)** Hysteresis loops of NdPd$_5$Ge$_3$ from -9 T to 9 T at different temperatures. **(Inset)** Field-dependence of magnetization from 0 to 3 T at 1.8 K and 2 K for NdPd$_5$Ge$_3$. The green straight lines are guide for the eye. **(d)** **(Main panel)** Expansion of the observed hysteresis loops for NdPd$_5$Ge$_3$ from -0.2 T to 0.2 T at low temperatures. **(Inset)** First derivative of magnetization curves of NdPd$_5$Ge$_3$ at 1.8 K, 2 K and 2.5 K between 1 and 3.5 T. The peak indicates the field where the magnetic-field-induced transition occurs.

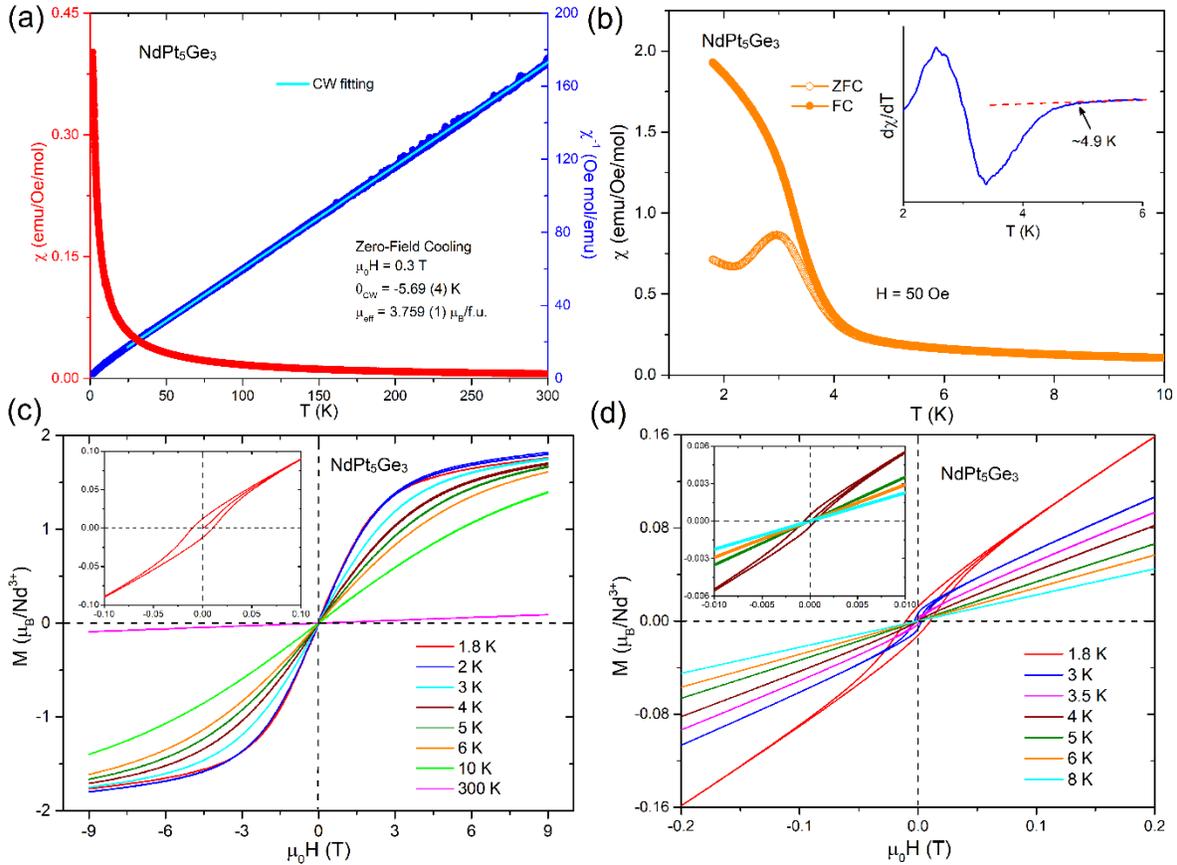

**Figure 4. (a)** Temperature-dependence of the magnetic susceptibility ($\chi$), and $1/\chi$ of NdPt$_5$Ge$_3$ between 1.8 K and 300 K under applied magnetic field of 0.3 T. measured in the ZFC mode. The cyan line represents the Curie-Weiss fitting. **(b) (Main panel)** Temperature-dependence of magnetic susceptibility ($\chi$) of NdPt$_5$Ge$_3$ between 1.8 K and 10 K under an applied magnetic field of 50 Oe (0.005 T). **(Inset)** First derivative of $\chi$ *vs* T obtained in the ZFC mode from 2 K to 6 K. **(c) (Main panel)** Hysteresis loops of NdPt$_5$Ge$_3$ from -9 T to 9 T at multiple temperatures. **(Inset)** Zoom-in of the hysteresis loop at 1.8 K between -0.1 T and 0.1 T. **(d) (Main panel)** Hysteresis loops of NdPt$_5$Ge$_3$ from -0.2 T to 0.2 T at low temperatures. **(Inset)** Zoom-ins of hysteresis loops at 4 K, 5 K, 6 K and 8 K from -0.01 T to 0.01 T.

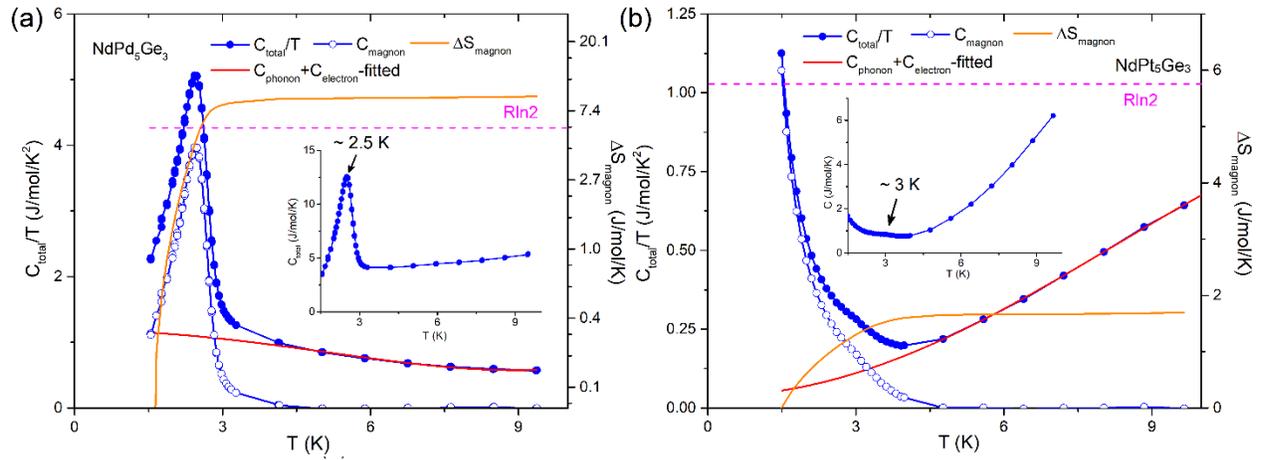

**Figure 5. (Main panel)** Heat capacity of **(a)** NdPd$_5$Ge$_3$; **(b)** NdPt$_5$Ge$_3$ from 1.5 K to 10 K. Solid blue circle with line represents C$_{total}$/T while empty blue circle with line stands for C$_{magnon}$/T. Red and orange solid lines indicate (C$_{phonon}$+C$_{electron}$)/T and entropy change from magnetic ordering ($\Delta$S$_{magnon}$). Dashed magenta line is the value of Rln2. The entropy change from magnetic ordering of NdPd$_5$Ge$_3$ was plotted natural logarithmically to make the figure clear. **(Inset)** Temperature-dependence of total heat capacity of **(a)** NdPd$_5$Ge$_3$; **(b)** NdPt$_5$Ge$_3$ from 1.5 K to 10 K.